\documentclass[runningheads]{llncs}
\usepackage[T1]{fontenc}
\usepackage[pdftex]{graphicx}
\usepackage{url}
\usepackage{amsmath}
\usepackage{float}
\begin{document}
\title{Food Recommendation With Balancing \\Comfort and Curiosity}
%
%
\author{Yuto Sakai\inst{1}\orcidID{0009-0006-1011-1427} \and
Qiang Ma\inst{2}\orcidID{0000-0003-3430-9244}
}
\authorrunning{Y. Sakai, Q. Ma}
%
\institute{Kyoto University, Japan \email{sakai.yuto.b66@kyoto-u.jp}\and
Kyoto Institute of Technology, Japan
\email{qiang@kit.ac.jp}}
\maketitle              
\begin{abstract}
Food is a key pleasure of traveling, but travelers face a trade-off between exploring curious new local food and choosing comfortable, familiar options.
This creates demand for personalized recommendation systems that balance these competing factors.
To the best of our knowledge, conventional recommendation methods cannot provide recommendations that offer both curiosity and comfort for food unknown to the user at a travel destination.
In this study, we propose new quantitative methods for estimating comfort and curiosity: Kernel Density Scoring (KDS) and Mahalanobis Distance Scoring (MDS).
KDS probabilistically estimates food history distribution using kernel density estimation, while MDS uses Mahalanobis distances between foods.
These methods score food based on how their representation vectors fit the estimated distributions.
We also propose a ranking method measuring the balance between comfort and curiosity based on taste and ingredients.
This balance is defined as curiosity (return) gained per unit of comfort (risk) in choosing a food.
For evaluation the proposed method, we newly collected a dataset containing user surveys on Japanese food and assessments of foreign food regarding comfort and curiosity.
Comparing our methods against the existing method, the Wilcoxon signed-rank test showed that when estimating comfort from taste and curiosity from ingredients, the MDS-based method outperformed the Baseline, while the KDS-based method showed no significant differences.
When estimating curiosity from taste and comfort from ingredients, both methods outperformed the Baseline.
The MDS-based method consistently outperformed KDS in ROC-AUC values.

\end{abstract}
\section{Introduction}
Trying new food or local specialties that one has never eaten before is one of the most enjoyable aspects of traveling, satisfying our curiosity, which is one of the most important motivations to travel.
It is desirable that the food that are eaten are made up of ingredients that one can eat and they taste good according to one's own taste.
Especially when it comes to food during travel, there is often no second chance to make a different choice if the food does not meet one's expectations.

Research on food recommendation has been actively conducted.
Gao et al. \cite{Gao2020} and Thongsri et al. \cite{Thongsri2022} proposed food recommendation systems that satisfy users' food preferences.
In the context of meals during travel, if recommendations are made using users' food preferences as in conventional methods, the recommended food are likely to be close to the user's preferred taste, ensuring a sense of comfort in enjoying the meal.
However, since the taste is similar to the foods they usually enjoy, it does not accompany curiosity.

Correia et al. \cite{Correia2019} concluded that the uniqueness and tradition of food are important factors in recommending food to tourists visiting a travel destination.
By recommending food based on the local food or popular food in that area, it is possible to recommend food that satisfy users' curiosity.
However, the recommendation results may include food made with ingredients that the user cannot eat or food with strong flavors that do not suit their palate.
Therefore, comfort is not guaranteed.

To the best of our knowledge, there is no method yet that recommends food at travel destinations that provide both comfort and curiosity for unknown food.
In our previous research \cite{YutoMa}, we proposed a neural network model that takes recipe data composed of food image data, ingredients, and cooking procedures as input and outputs the taste representation vector and ingredient representation vector.
We demonstrated through case studies that it is possible to discover food that ensure comfort and curiosity by comparing the representation vectors of the target food's taste and ingredients with the user's past food history using interaction data from recipe sites.
However, how to recommend food with balancing comfort and curiosity is still a challenge.

În this study, we propose the concepts of comfort and curiosity in food recommendation and their quantification methods, KDS (Kernel Density Scoring) and MDS (Mahalanobis Distance Scoring), to recommend food that balance comfort and curiosity.
The user's food history distribution is probabilistically estimated using kernel density estimation in KDS and distance-based estimation using Mahalanobis distances between foods in MDS.
We evaluate how well the representation vectors of the recommended food fit the estimated distribution.
The comfort-curiosity score indicates greater comfort with smaller values and greater curiosity with larger values.
Additionally, we propose a ranking measure that estimates the balance of comfort and curiosity from the taste and ingredients of the food.
By ranking candidate foods based on this measure, we realize food recommendation.
The balance of comfort and curiosity is defined by associating the taste and ingredients of the food with the concepts of risk and return on a one-to-one basis, formulating curiosity as the return obtained per unit of risk in choosing a food.

\begin{figure}[!tb]
    \begin{center}
      \includegraphics[width=0.6\linewidth]{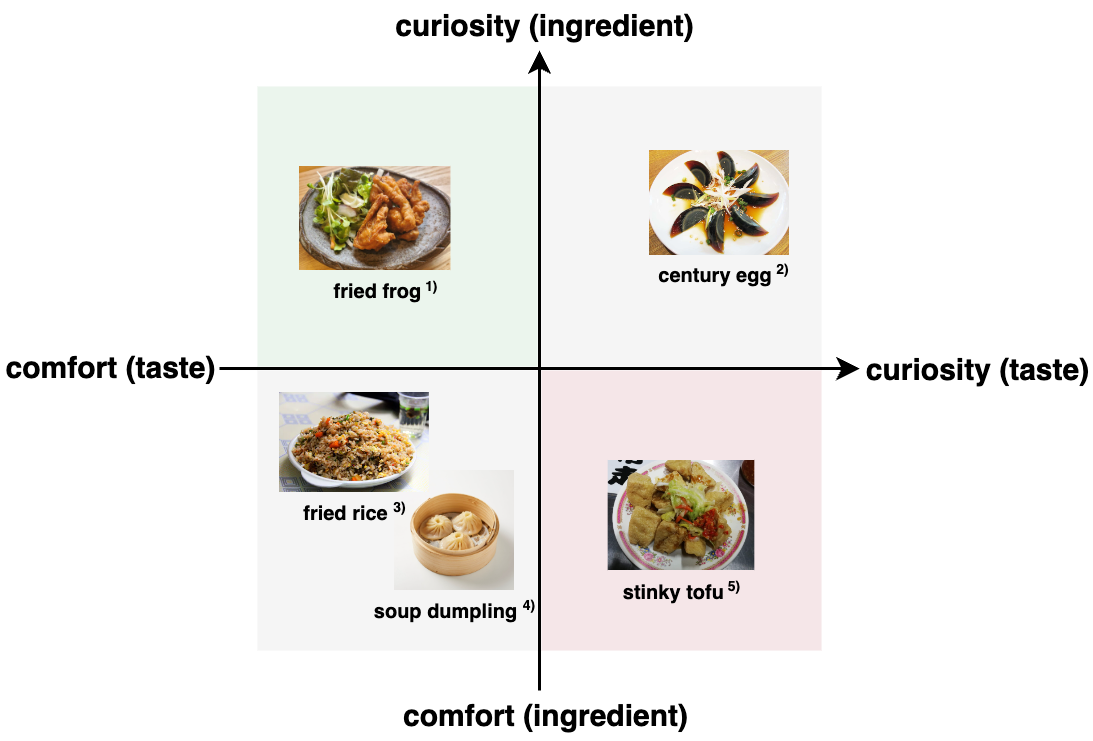}
      \caption{Example of the recommended food}
      \label{recommendation_example}
    \end{center}
\end{figure}

Figure \ref{recommendation_example} shows an example of the proposed method and recommended foods.
It assumes a user who regularly eats familiar food in Japan and visits China for the first time.
In this case, conventional preference-based recommendation methods would recommend food like fried rice or soup dumpling, which are commonly eaten in daily life.
Also, recommending food like century eggs, which are challenging in both taste and ingredients for the user, is not desirable.
In this study, we achieve a balance of comfort and curiosity by recommending food like stinky tofu, which arouses curiosity in taste but is comforting in familiar ingredients, or conversely, food like fried frog, which are close in taste to food the user has eaten before but have unknown ingredients that arouse curiosity.

\footnotetext[1]{\url{https://x.gd/ASuZB}}
\footnotetext[2]{\url{https://x.gd/yGA7a}}
\footnotetext[3]{\url{https://x.gd/krZJr}}
\footnotetext[4]{\url{https://x.gd/euQxs}}
\footnotetext[5]{\url{https://x.gd/8UesA}}

The major contributions of this study are as follows:
\begin{itemize}
  \item We propose a novel recommendation method considering both comfort and curiosity. At first, we define and quantifies comfort and curiosity from aspects of taste and ingredients. Then, we propose an integrated measure to estimate the balance of comfort and curiosity to food recommendation. (Section \ref{proposed_method})
  \item We construct a new dataset that records user emotions towards food, evaluated from the perspectives of taste and ingredients, based on interaction data between users and food collected through crowdsourcing. (Section \ref{experiment})
  \item We demonstrate through experiments that the proposed recommendation method can achieve food recommendations that ensure both comfort and curiosity. (Section \ref{experiment})
\end{itemize}

\section{Related Work}
\label{related_work}
\subsection{Research on Relevance-Oriented Recommendations}
Conventional food recommendation systems has focused on recommending food based on the relevance of historical data, specifically the user's food preferences.
Gao et al. \cite{Gao2020} define food recommendation as the problem of predicting a user's preference for a recipe.
They propose a recommendation system based on a neural network model called Hierarchical Attention based Food Recommendation (HAFR) to estimate food preferences.
HAFR estimates the probability that a user accepts a particular recipe using three inputs: historical data indicating whether the user has eaten a specific recipe, ingredient data, and image data of the food.
Ueda et al. \cite{Ueda2014} also propose a recipe recommendation method based on food preferences.
Their research estimates a user's food preferences, including liked and disliked ingredients, from recipes the user has viewed or cooked in the past, and recommends recipes with higher scores when they contain more preferred ingredients.

Thongsri et al. \cite{Thongsri2022} propose a method that not only estimates food that match a user's preferences using collaborative filtering but also calculates basal metabolic rate based on features such as gender, height, and weight, solving a knapsack problem with calorie constraints to recommend food that align with user preferences while maintaining health.

Zhang et al. \cite{Zhang2020} propose a personalized restaurant recommendation system that extracts visual features from images using deep convolutional neural networks and combines them with collaborative filtering methods.

\subsection{Research on Serendipity-Oriented Recommendations}
Research on recommendation systems focus on modeling user preferences and recommending items that match those preferences.
However, merely recommending items related to the user's past behavior can lead to feelings of boredom or dissatisfaction.
There is a risk of missing out on items that offer unexpected value.
In recent trends, research on serendipity-oriented recommendations has become active to address these issues.
The concept of serendipity in recommendation systems has not yet reached a consensus on its definition, but Fu et al. \cite{Fu2023} argue that serendipity consists of four elements: unexpectedness, novelty, diversity, and relevance.

Li et al. \cite{Li2020} propose a recommendation system that incorporates unexpectedness as an element to provide surprise and satisfaction.
Their proposed method embeds items experienced by the user in the past into a latent space and forms interest clusters using the mean shift clustering method.
They model unexpectedness by calculating the weighted distance between recommended items and each cluster.
Lo et al. \cite{Lo2021} incorporate the concept of novelty into recommendations by focusing on items that have existed for a long time but are unpopular.
Cui et al. \cite{Cui2020} introduce diversity into recommendations by defining the similarity between items using Pearson's correlation coefficient and determining the diversity of an item as the reciprocal of the sum of similarities with all items in the user's history set.
Zhang et al. \cite{Zhang2012} demonstrate that it is possible to simultaneously improve unexpectedness, novelty, and diversity in music recommendations.
Their research proposes entropy-based and graph-based algorithms in addition to conventional accuracy-focused recommendation frameworks, achieving serendipity-oriented recommendations by linearly complementing recommendation rankings output by multiple recommendation frameworks.

Recommendations based on user food preferences are useful in everyday situations, but they may not suffice for food at travel destinations in unfamiliar lands, where there is a desire to try local and unusual food.
Therefore, research is being conducted on recommendation factors beyond relevance.
Kauppinen et al. \cite{Kauppinen2013} conducted interviews to understand memorable dining experiences. They identified four main themes, including tourism, and emphasized tourists' willingness to try new meals as a key characteristic of their dining experiences.
Jiménez et al. \cite{Jimenez2016} conducted research analyzing the relationship between gastronomy, culture, and tourism, revealing that tourists consider food an important part of the cultural identity of tourist destinations.
Correia et al. \cite{Correia2019} analyzed data from tourists in Hong Kong, highlighting food quality, uniqueness, tradition, and service as key recommendation factors, with a focus on uniqueness and tradition.

In short, There is no agreed-on definition of serendipity in recommendation system research, and each study independently defines it while associating it with concepts such as unexpectedness, novelty, and diversity.
Similarly, the concepts of comfort and curiosity addressed in this study have not yet been defined in the context of recommendation systems.
To the best of our knowledge, recommendation methods that balance these two concepts does not yet exist.

\section{Proposed Method}
\label{proposed_method}
\begin{figure}[!tb]
    \centering
    \includegraphics[width=0.47\linewidth]{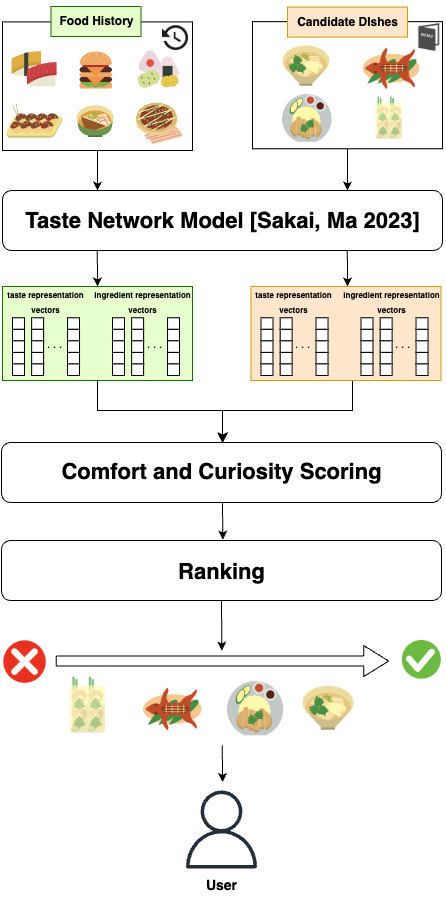}
    \caption{Overview of Our Recommendation System}
    \label{recommendation_system_architecture}
\end{figure}

The proposed recommendation system is illustrated in Figure \ref{recommendation_system_architecture}.
First, it receives the user's food history and information on candidate food for recommendation, and outputs a set of representation vectors for each food using the taste network model proposed in our previous research \cite{YutoMa}.
Next, the comfort and curiosity scoring module inputs the set of representation vectors of the history and candidate foods, and calculates the comfort and curiosity scores for each candidate food.
Finally, the ranking module measures the balance of comfort and curiosity of candidate foods and ranks.
This section details the newly defined quantitative analysis methods for comfort and curiosity, the measure for estimating the balance of comfort and curiosity, and the recommendation method for food that incorporate both comfort and curiosity.
The definitions of the terms used in the explanation are shown in Table \ref{term_description}.

\begin{table}[!tb]
  \centering
  \caption{Definition of Terms in the Proposed Method}
  \label{term_description}
  \begin{tabular}{l|l}
  \hline \hline
  $F$ & Set of candidate foods for recommendation \\
  $H$ & Set of foods eaten in the past \\
  $p(\cdot)$ & Probability density function for the distribution of history $H$ \\
  $Score_{f}$ & Score for food $f$ \\
  $Score'_{f}$ & Scaled score \\
  $Score^{taste}$ & Score for taste representation vector \\
  $Score^{ing}$ & Score for ingredient representation vector \\
  $d_{i,j}$ & Distance between $h_i, h_j \in H$ \\
  $d_i$ & Distance between $h_i \in H, f \in F$ \\
  $d_{in}$ & Average distance between all pairs of foods within $H$ \\
  $d_{out}$ & Average distance between candidate food and all foods within $H$ \\
  \hline \hline
  \end{tabular}
\end{table}

\subsection{Definition of Comfort and Curiosity}
\label{definition_comfort_curiosity}
In this study, ``comfort'' refers to the feeling of reassurance when selecting a food that aligns with one's taste and ingredient preferences.
On the other hand, ``curiosity'' refers to the desire to try new or region-specific food that one does not usually eat.
Therefore, a food that possesses comfort is defined as one that belongs to the domain formed by the taste and ingredient trends of foods previously eaten.
Additionally, a food that possesses curiosity is defined as one that differs from the taste and ingredient trends of foods previously eaten.

\subsection{Quantification of Comfort and Curiosity}
\label{scoring_comfort_curiosity}
This section describes the methods for quantifying comfort and curiosity as defined in Section \ref{definition_comfort_curiosity}.
As mentioned before, we propose two methods: Kernel Density Scoring (KDS) and Mahalanobis Distance Scoring (MDS).

\paragraph{Kernel Density Scoring (KDS)}
The score $Score_f$ for a food $f$ is evaluated using a probabilistic approach based on how close the representation vector of food $f$ is to the distribution formed by the representation vectors of food $h$ included in the set of foods $H$ previously eaten.
Since the distribution of the food history may not fit a known distribution, a non-parametric method, Kernel Density Estimation (KDE)\cite{KernelDensityEstimation}, is used to estimate the probability density function of the distribution.
The procedure is as follows:
\begin{enumerate}
  \item Perform principal component analysis on the set of representation vectors of food $h_1, h_2, ..., h_n \in H$ and the candidate food $f$, reducing each representation vector to a two-dimensional vector.
  \item Estimate the probability density function $p(h)$ using KDE on the reduced-dimensional representation vectors of food $h_1, h_2, ..., h_n \in H$.
  \item Calculate the probability density $p(f)$ that food $f$ belongs to the distribution of history $H$.
  \item Take the negative logarithm of the probability density obtained in step 3, and set $Score_{f} = -\log p(f)$.
\end{enumerate}

\medskip
\paragraph{Mahalanobis Distance Scoring (MDS)}
The score $Score_{f}$ for a food $f$ is evaluated using a distance-based approach based on how close the representation vector of food $f$ is to the distribution formed by the representation vectors of food $h$ included in the set of foods $H$ previously eaten.
The spatial relationship between the candidate food and the history distribution is captured while considering the internal structure formed by the history.
The Mahalanobis distance is adopted as the distance measure to consider the spread of the distribution.
The procedure is as follows:
\begin{enumerate}
  \item Perform principal component analysis on the set of representation vectors of foods $h_1, h_2, ..., h_n \in H$ and the candidate food $f$, reducing each representation vector to a two-dimensional vector.
  \item Calculate the average Mahalanobis distance $d_{in}$ between all pairs of reduced-dimensional representation vectors of food $h_1, h_2, ..., h_n \in H$.
  \begin{align*}
    d_{in} &= \cfrac{1}{n(n-1)}\sum_{i=1}^{n}\sum_{j=1, j\neq i}^{n}d_{i,j}
  \end{align*}
  \item Calculate the average Mahalanobis distance $d_{out}$ between the representation vector of food $f$ and each representation vector of food $h_i$ included in the history.
  \begin{align*}
    d_{out} &= \cfrac{1}{n}\sum_{i=1}^{n}d_i
  \end{align*}
  \item Take the ratio of the values obtained in steps 2 and 3, and set $Score_{f} = \cfrac{d_{out}}{d_{in}}$.
\end{enumerate}

\subsection{Ranking Method}
\label{ranking_method}
This section describes the method for ranking the set of recommended foods using the score $Score_{f}$ defined in Section \ref{scoring_comfort_curiosity}.

First, the scaling of $Score_{f}$ is performed.
When the score for a food $h_i$ included in the history $H$ is $Score_{h_i}$, the scaled value $Score'_{h_i}$ is defined as follows:
\begin{align*}
  Score_{max} &= \max_{h \in H}Score_{h} \\
  Score'_{h_i} &= \cfrac{Score_{h_i}}{Score_{max}} \\
\end{align*}
Similarly, the scaled score $Score'_{f_i}$ for a food $f_i$ included in the set of recommended foods $F$ is defined as follows:
\begin{align*}
  Score'_{f_i} &= \cfrac{Score_{f_i}}{Score_{max}} \\
\end{align*}

Next, the degree of balance between comfort and curiosity is calculated.
Comfort and curiosity are obtained from the taste and ingredients of the food, respectively, and the balance $Score^{total}_{f_i}$ for each food $f_i \in F$ is defined as follows.
Here, the formula is shown for the case where comfort is obtained from the taste and curiosity from the ingredients.
\begin{equation}
\label{score}
  Score^{total}_{f} = \cfrac{Score'^{ing}_{f}-r_f}{Score'^{taste}_{f}}
\end{equation}

$r_f$ is defined as following procedures.
First, when calculating the score \\$Score^{taste}$, we identify the food $h_{min}$ with the smallest value of $Score^{taste}_{h_i}$.
The food $h_{min}$, which has the smallest value of $Score^{taste}$, is considered to be a food with assured comfort as it best fits the distribution of foods previously eaten.
Then, we calculate the curiosity score $Score^{ing}_{h_{min}}$ for food $h_{min}$, scale it, and set $r_f = Score'^{ing}_{h_{min}}$.

Finally, for all foods $f_i \in F$ included in the set of recommended food candidates, we calculate $Score^{total}_{f_i}$ using the formula (\ref{score}), and rank the foods by sorting the values in descending order to achieve a balance between comfort and curiosity.

\section{Experiment}
\label{experiment}

\subsection{Dataset}
\label{experiment_dataset}
In this study, the task is to recommend food that balance comfort and curiosity.
To the best of out knowledge, there is no existing research addressing a similar task, nor is there a dataset that records evaluations of comfort and curiosity.
Therefore, we conducted a survey using crowdsourcing to construct a dataset that records real users' food histories and their evaluations of comfort and curiosity towards food.
The platform used for crowdsourcing was ``Lancers''\footnote[6]{\url{https://www.lancers.jp/}}.
Workers were informed that their response data would be used for research and analysis, and consent was obtained before collecting responses.

The crowdsourcing task consisted of questions regarding a set of foods, including 12 types of Japanese food and 9 or 10 types of foods from either Southeast Asia, China, or Europe.
Workers answered the following four questions for each food:
{\renewcommand{\theenumi}{\roman{enumi}}
\begin{enumerate}
  \item Whether they have eaten this food before.
  \item Whether they feel this food suits their taste and whether they would like to try it.
  \item Whether they feel this food suits their preference for ingredients and whether they would like to try it.
  \item List foods they usually eat in their daily life, other than those listed in the survey.
\end{enumerate}}
For question i, workers answered with either ``yes'' or ``no.''
For questions ii and iii, they selected from the following four options:
\begin{enumerate}
  \item Feels suitable + Would like to try
  \item Feels suitable + Would not like to try
  \item Feels unsuitable and anxious + Would like to try
  \item Feels unsuitable and anxious + Would not like to try
\end{enumerate}
For question iv, workers provided as many foods as possible that they usually eat in their daily life in a free-text format.

These tasks were created for each set of foods composed of Japanese and Southeast Asian food, Japanese and Chinese food, and Japanese and European food, with 100 different workers responding to each task.

From the collected response data, the following four types of datasets were created:
\paragraph{\textbf{\textit{all\_food}}}
A dataset that records the recipe data of 40 types of foods prepared as questions in the crowdsourcing task.
The breakdown of the 40 foods is 12 types of Japanese food, 10 types of Southeast Asian food, 9 types of Chinese food, and 9 types of European food. 
Each recipe contains information such as food ID, food name, ingredients, cooking instructions, and food image.

\paragraph{\textbf{\textit{extended\_food}}}
A dataset created by extracting food names included in the food history answered in question iv of the crowdsourcing and collecting recipes that can be made.
It records 170 types of food recipes, and the information each recipe contains is the same as in the \textit{all\_food} dataset.

\paragraph{\textbf{\textit{user\_food\_interaction}}}
A dataset that records the interaction between workers and foods, as well as the evaluation values of comfort and curiosity, answered in questions i, ii, and iii of the crowdsourcing.
The minimum components of the dataset include worker ID, food ID from \textit{all\_food}, experience (answer to question i), taste evaluation (answer to question ii), and ingredient evaluation (answer to question iii).

\paragraph{\textbf{\textit{user\_food\_extended\_interaction}}}
A dataset that records the interaction between workers and foods answered in question iv of the crowdsourcing.
The minimum components of the dataset include worker ID and food ID from \textit{extended\_food}.

\subsection{Experiment Settings}
\label{experiment_setting}
We evaluate the performance of the proposed recommendation method using the dataset described in Section \ref{experiment_dataset}.
The experimental procedure is as follows:
\begin{enumerate}
  \item For users recorded in the dataset, set all foods from the region targeted in the crowdsourcing task as candidate foods for recommendation, and obtain a ranking based on the balance of comfort and curiosity defined in Section \ref{ranking_method}.
  The user's food history includes all foods they have eaten, as recorded in the \textit{user\_food\_interaction} and \textit{user\_food\_extended\_interaction} datasets.
  \item Define the relevant items for recommendation to users in Experiment 1, where comfort is derived from taste and curiosity from ingredients, and Experiment 2, where curiosity is derived from taste and comfort from ingredients.
  \begin{itemize}
    \item The relevant items for Experiment 1 are the set of foods with a taste evaluation of ``feels suitable'' and an ingredient evaluation of ``would like to try.'' This corresponds to the green area in the second quadrant of Figure \ref{recommendation_example}.
    \item The relevant items for Experiment 2 are the set of foods with a taste evaluation of ``would like to try'' and an ingredient evaluation of ``feels suitable.'' This corresponds to the red area in the fourth quadrant of Figure \ref{recommendation_example}.
  \end{itemize}
  \item Compare the output ranking with the set of correct items to evaluate the performance of the recommendation.
\end{enumerate}

The ranking performance is evaluated by using $Precision@K, Recall@K,$ and $NDCG@K(K=1, 3, 5)$.
The recommendation methods evaluated include two methods using MDS and KDS for quantifying comfort and curiosity, and the Baseline method that generates rankings randomly, making a total of three methods.
The evaluation value of the Baseline method is obtained by repeating the evaluation of the ranking output by randomly arranging the candidate food set $10^6$ times and using the average value.

Finally, based on the evaluation values of each method, we statistically test the effectiveness of the proposed recommendation method.
In this test, considering that the set of values for each evaluation metric does not guarantee a normal distribution, we perform a two-sided test using the Wilcoxon signed-rank test.
The data used for the test includes a total of 27 values based on the combination of three evaluation metrics ($Precision, Recall, NDCG$), three $K$ values, and three regions (Southeast Asia, China, Europe) for each method.
Then, with a significance level of $0.05$, we test whether there is a significant difference between the evaluation results of the Baseline and the proposed method.


\subsection{Experiment Results}
\label{experiment_result}
\subsubsection{Experiment 1}
The recommendation performance when obtaining comfort from taste and curiosity from ingredients is shown in Tables \ref{taste_comfort_ranking_southeast_asia}, \ref{taste_comfort_ranking_china}, \ref{taste_comfort_ranking_europe}, and \ref{taste_comfort_ranking_all_regions}.
The Wilcoxon signed-rank test results showed that the $p$-value between the evaluation values of the MDS-based method and the Baseline was $0.01523 < 0.05$, confirming that the MDS-based method is superior to the Baseline.
On the other hand, the $p$-value between the evaluation values of the KDS-based method and the Baseline was $0.1023 > 0.05$, indicating no significant difference in performance.

\begin{figure*}[!tb]
  \centering
  \begin{minipage}{0.49\linewidth}
    \includegraphics[width=0.9\linewidth]{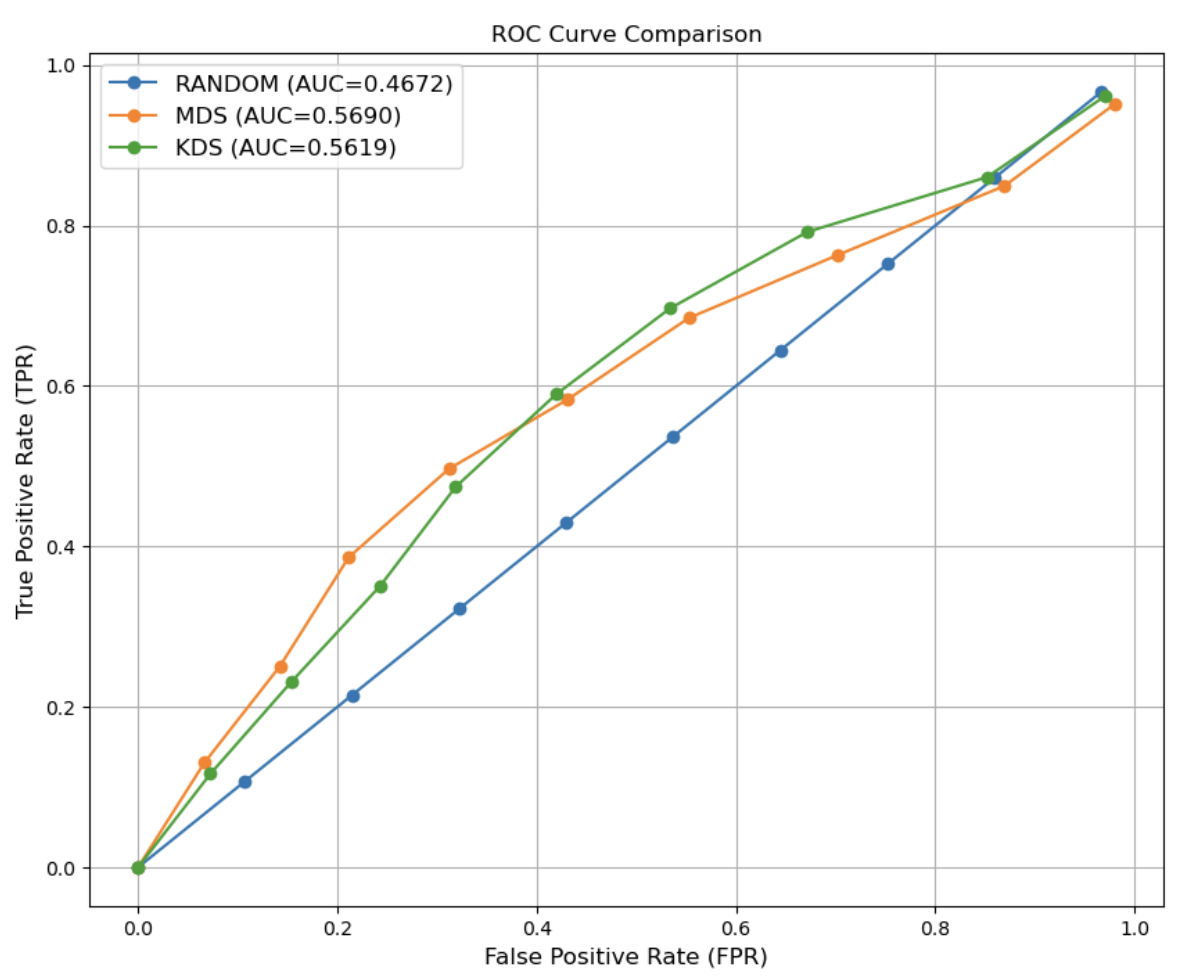}
  \end{minipage}
  \begin{minipage}{0.49\linewidth}
    \includegraphics[width=0.9\linewidth]{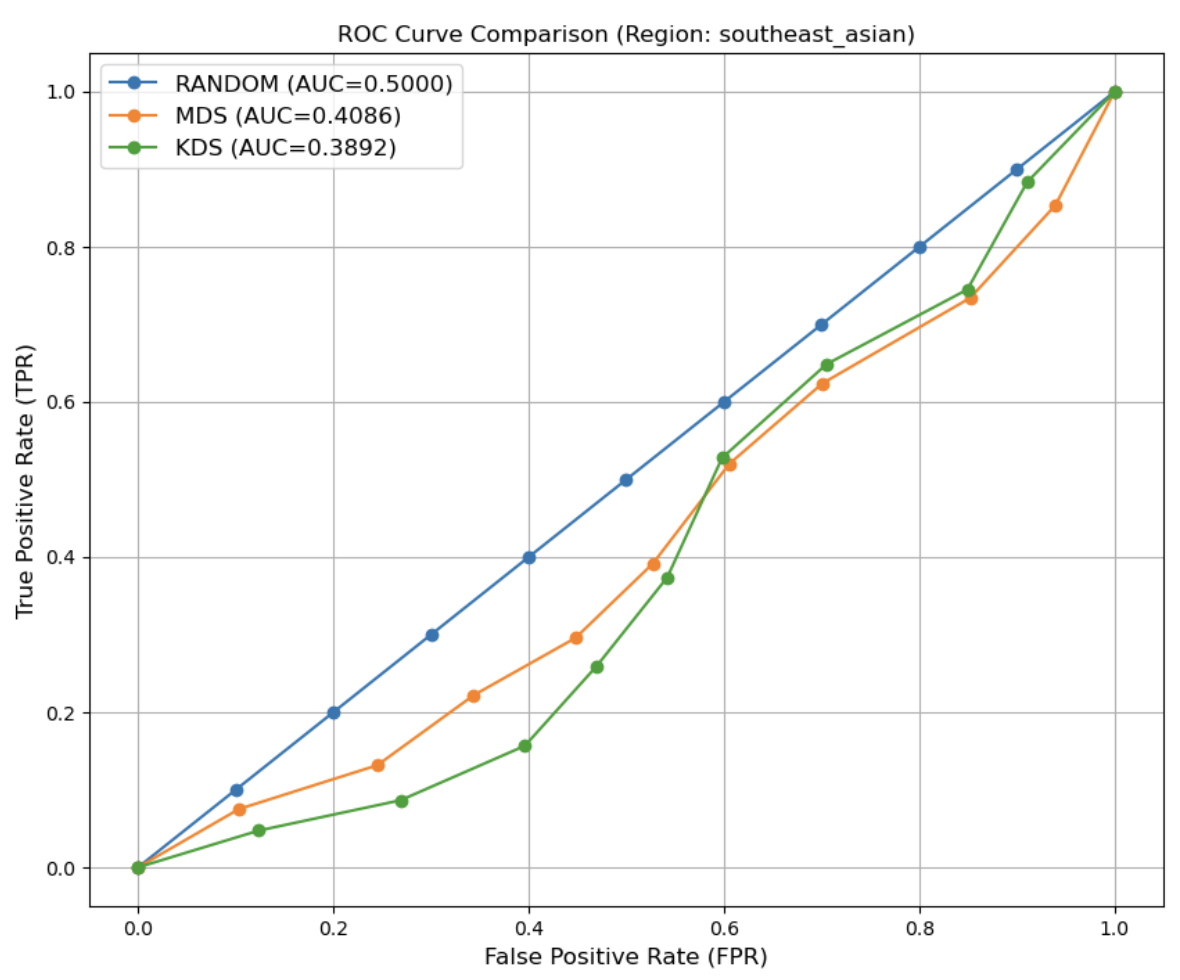}
  \end{minipage}
  \begin{minipage}{0.49\linewidth}
    \includegraphics[width=0.9\linewidth]{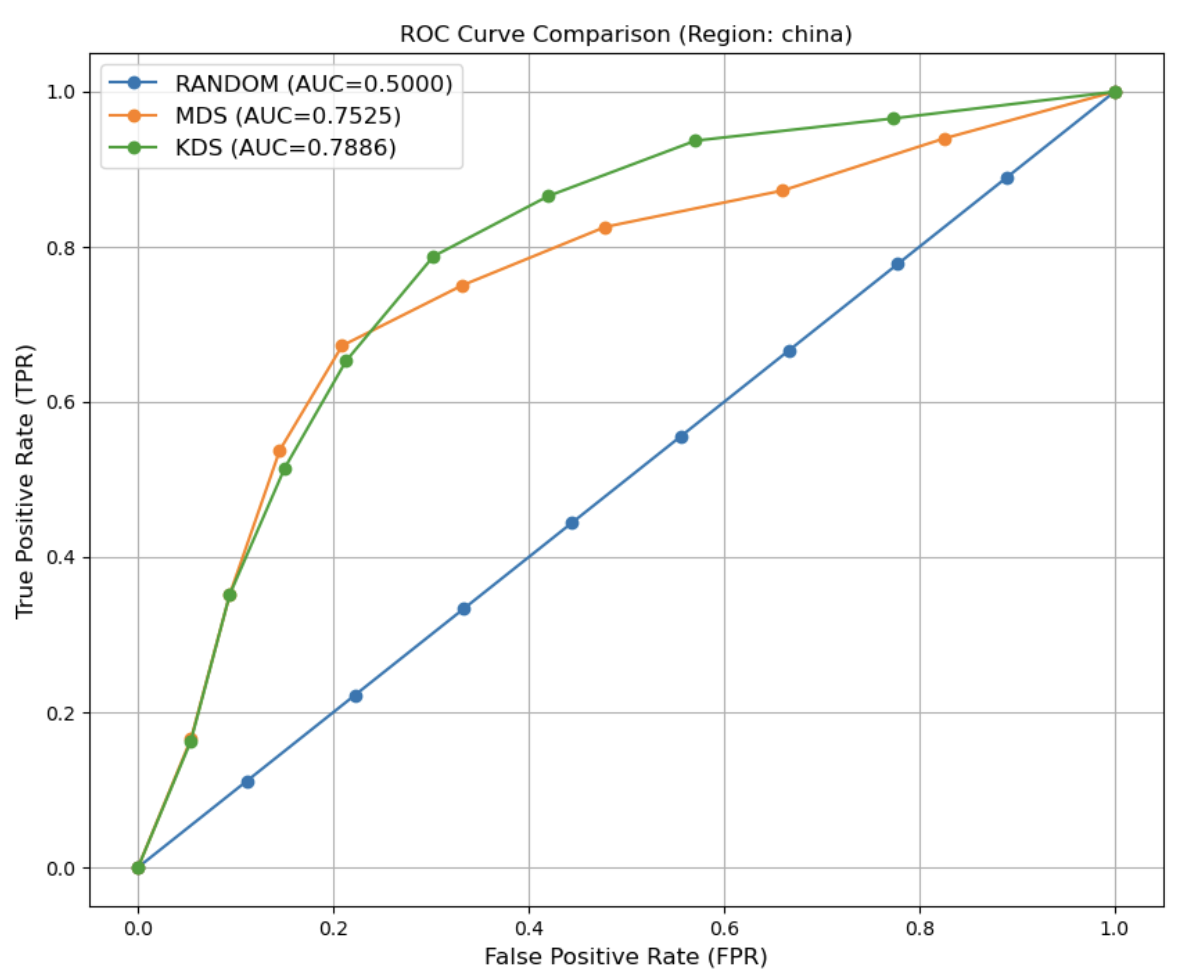}
  \end{minipage}
  \begin{minipage}{0.49\linewidth}
    \includegraphics[width=0.9\linewidth]{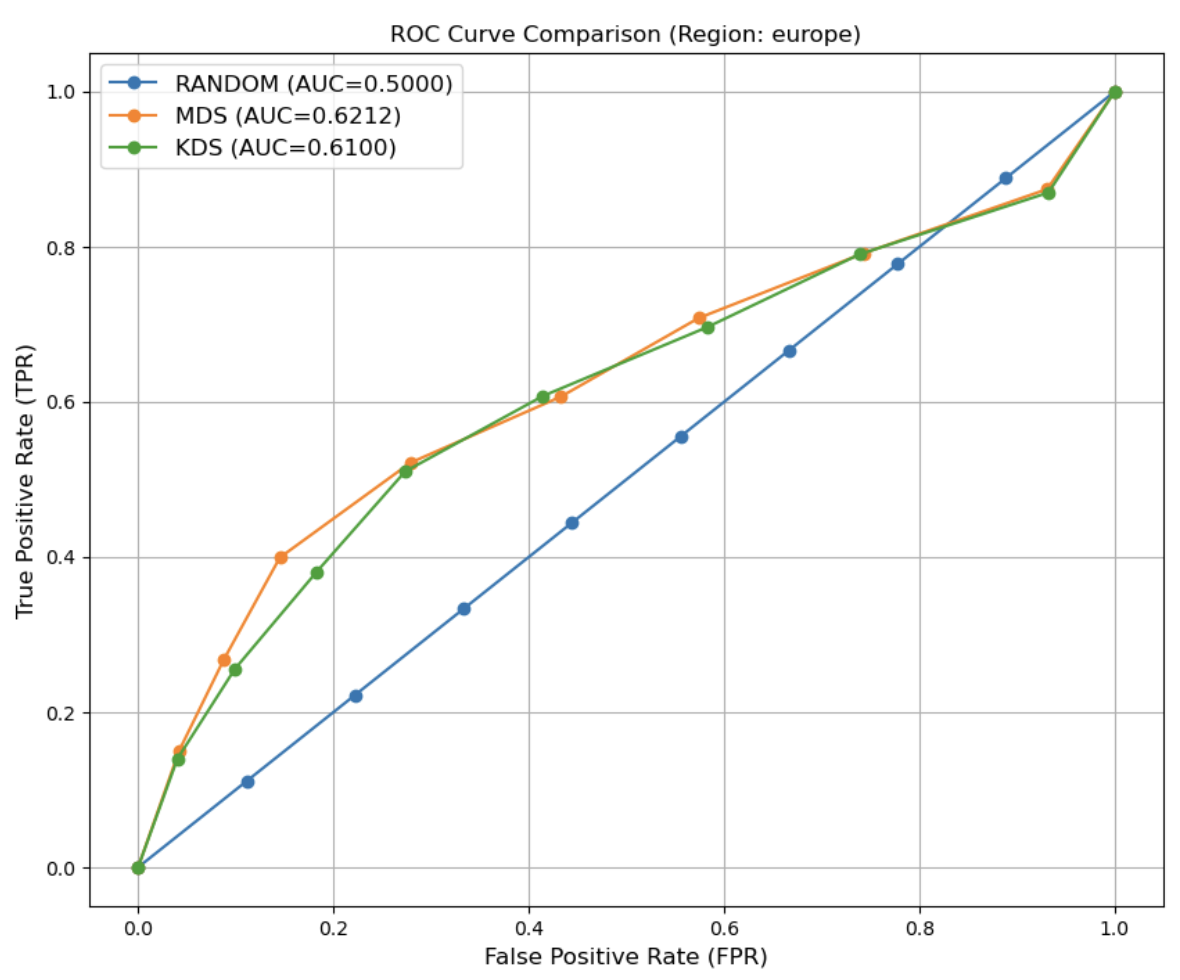}
  \end{minipage}
  \caption{ROC Curve: Taste-Comfort (Top Left: Average Across All Regions, Top Right: Southeast Asian Food, Bottom Left: Chinese Food, Bottom Right: European Food)}
  \label{roc_curve_taste_comfort}
\end{figure*}

Figure \ref{roc_curve_taste_comfort} shows the ROC curve for taste-comfort.
Looking at the average ROC curve across all regions, the ROC curve of the proposed method based on MDS and KDS is positioned above the Baseline, indicating that the proposed method has higher recommendation performance.
In the comparison between the MDS-based and KDS-based methods, the AUC values were $0.5690$ for the former and $0.5619$ for the latter, demonstrating that the MDS-based method is superior to the KDS-based method.

Looking at the ROC curves for each region, Table \ref{taste_comfort_ranking_southeast_asia} shows that in the recommendation case for Southeast Asian food, both proposed methods perform worse than the Baseline.
The ROC curves of the KDS-based and MDS-based methods are positioned below that of the Baseline, indicating that the proposed methods are making recommendations that reverse positive and negative examples.
In the recommendation cases for Chinese and European food, the proposed methods show higher performance than the Baseline.
In particular, the recommendation for Chinese food is shown to be successful with high accuracy.

\vspace{-5mm}
\begin{table}[H]
  \centering
  \caption{Recommendation Performance of Southeast Asian Food}
  \label{taste_comfort_ranking_southeast_asia}
  \scalebox{0.9}{
    \begin{tabular}{l|llllllllll}
    \hline \hline
    Method & P@1 & P@3 & P@5 & R@1 & R@3 & R@5 & NDCG@1 & NDCG@3 & NDCG@5 \\\hline
    Baseline & {\bf 0.4753} & {\bf 0.4753} & {\bf 0.4753} & {\bf 0.1000} & {\bf 0.3000} & {\bf 0.5000} & {\bf 0.4753} & {\bf 0.4963} & {\bf 0.5406} \\
    MDS & 0.3951 & 0.3868 & 0.4123 & 0.0751 & 0.2220 & 0.3916 & 0.3951 & 0.3976 & 0.4418 \\
    KDS & 0.3210 & 0.3128 & 0.4049 & 0.0474 & 0.1567 & 0.3737 & 0.3210 & 0.3156 & 0.4023 \\\hline \hline
    \end{tabular}
  }
\end{table}

\vspace{-10mm}
\begin{table}[H]
  \centering
  \caption{Recommendation Performance of Chinese Food}
  \label{taste_comfort_ranking_china}
  \scalebox{0.9}{
    \begin{tabular}{l|lllllllll}
    \hline \hline
    Method & P@1 & P@3 & P@5 & R@1 & R@3 & R@5 & NDCG@1 & NDCG@3 & NDCG@5 \\\hline
    Baseline & 0.5040 & 0.5040 & 0.5040 & 0.1111 & 0.3333 & 0.5555 & 0.5040 & 0.5202 & 0.5619 \\
    MDS & {\bf 0.7216} & {\bf 0.7491} & 0.6619 & {\bf 0.1656} & {\bf 0.5378} & 0.7503 & {\bf 0.7216} & {\bf 0.7784} & 0.7782 \\
    KDS & {\bf 0.7216} & 0.7388 & {\bf 0.6887} & 0.1623 & 0.5152 & {\bf 0.7878} & {\bf 0.7216} & 0.7647 & {\bf 0.7967} \\\hline \hline
    \end{tabular}
  }
\end{table}

\begin{table}[H]
  \centering
  \caption{Recommendation Performance of European Food}
  \label{taste_comfort_ranking_europe}
  \scalebox{0.9}{
    \begin{tabular}{l|llllllllll}
    \hline \hline
    Method & P@1 & P@3 & P@5 & R@1 & R@3 & R@5 & NDCG@1 & NDCG@3 & NDCG@5 \\\hline
    Baseline & 0.6756 & 0.6756 & 0.6756 & 0.1111 & 0.3333 & 0.5556 & 0.6756 & 0.6846 & 0.7067 \\
    MDS & {\bf 0.8506} & {\bf 0.7969} & 0.7195 & {\bf 0.1507 }& {\bf 0.3994} & 0.6069 & {\bf 0.8506} & {\bf 0.8157} & {\bf 0.7929} \\
    KDS & 0.8161 & 0.7663 & {\bf 0.7241} & 0.1399 & 0.3801 & {\bf 0.6077} & 0.8161 & 0.7842 & 0.7842 \\\hline \hline
    \end{tabular}
  }
\end{table}

\vspace{-10mm}
\begin{table}[H]
  \centering
  \caption{Average Recommendation Performance Across Three Regions}
  \label{taste_comfort_ranking_all_regions}
  \scalebox{0.9}{
    \begin{tabular}{l|llllllllll}
    \hline \hline
    Method & P@1 & P@3 & P@5 & R@1 & R@3 & R@5 & NDCG@1 & NDCG@3 & NDCG@5 \\\hline
    Baseline & 0.5516 & 0.5516 & 0.5516 & 0.1074 & 0.3222 & 0.5371 & 0.5516 & 0.5670 & 0.6031 \\
    MDS & {\bf 0.6558} & {\bf 0.6443} & 0.5979 & {\bf 0.1305} & {\bf 0.3864} & 0.5829 & {\bf 0.6558} & {\bf 0.6639} & {\bf 0.6710} \\
    KDS & 0.6196 & 0.6060 & {\bf 0.6059} & 0.1165 & 0.3506 & {\bf 0.5897} & 0.6196 & 0.6215 & 0.6610 \\\hline \hline
    \end{tabular}
  }
\end{table}

\subsubsection{Experiment 2}
The recommendation performance when obtaining curiosity from taste and comfort from ingredients is shown in Tables \ref{taste_curiosity_ranking_southeast_asia}, \ref{taste_curiosity_ranking_china}, \ref{taste_curiosity_ranking_europe}, and \ref{taste_curiosity_ranking_all_regions}.
The Wilcoxon signed-rank test results showed that the $p$-value between the evaluation values of the MDS-based method and the Baseline was $0.00003978 < 0.05$, and the $p$-value between the evaluation values of the KDS-based method and the Baseline was $0.001938 < 0.05$.
This confirms that the proposed methods are superior to the Baseline.

\begin{figure*}[!tb]
  \centering
  \begin{minipage}{0.49\linewidth}
    \includegraphics[width=0.9\linewidth]{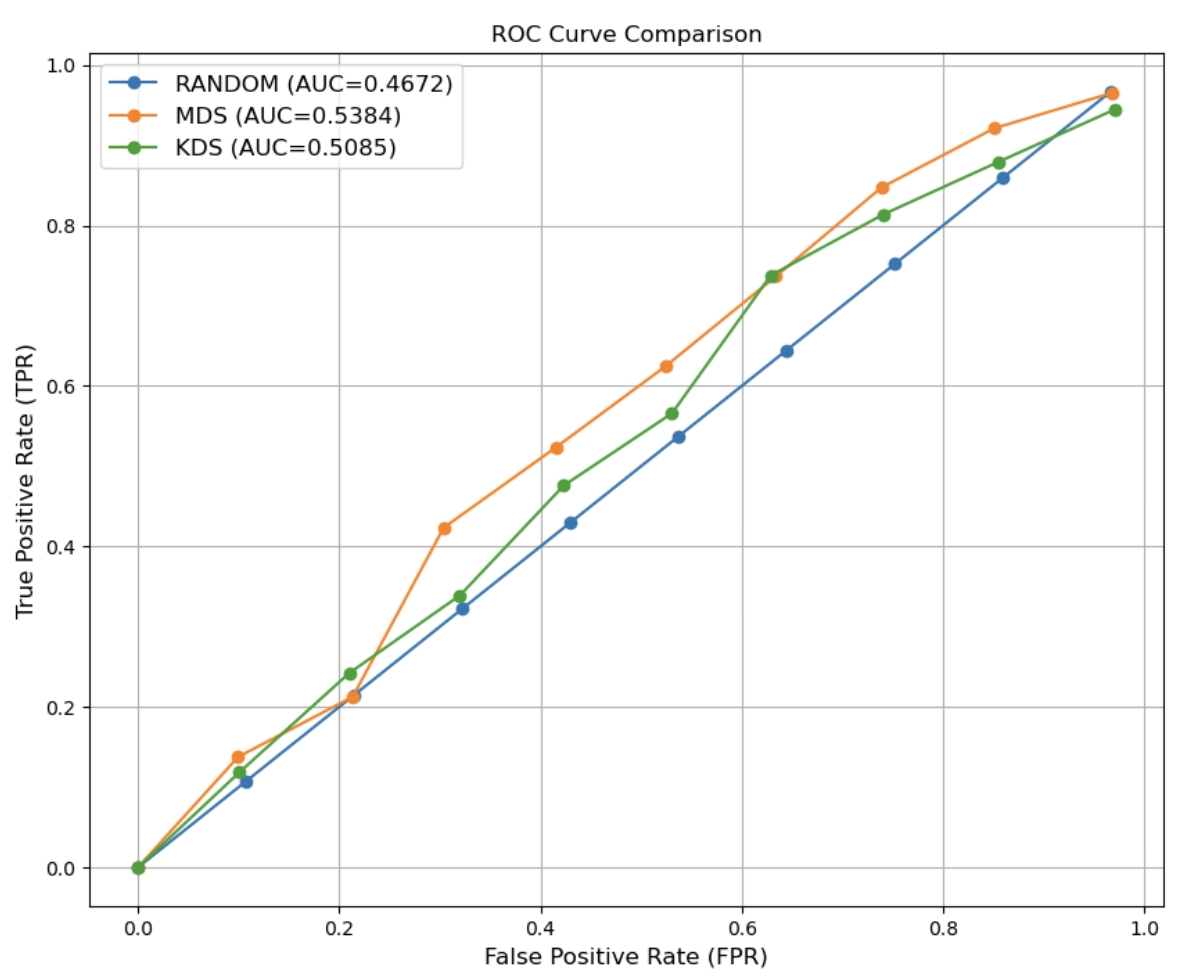}
  \end{minipage}
  \begin{minipage}{0.49\linewidth}
    \includegraphics[width=0.9\linewidth]{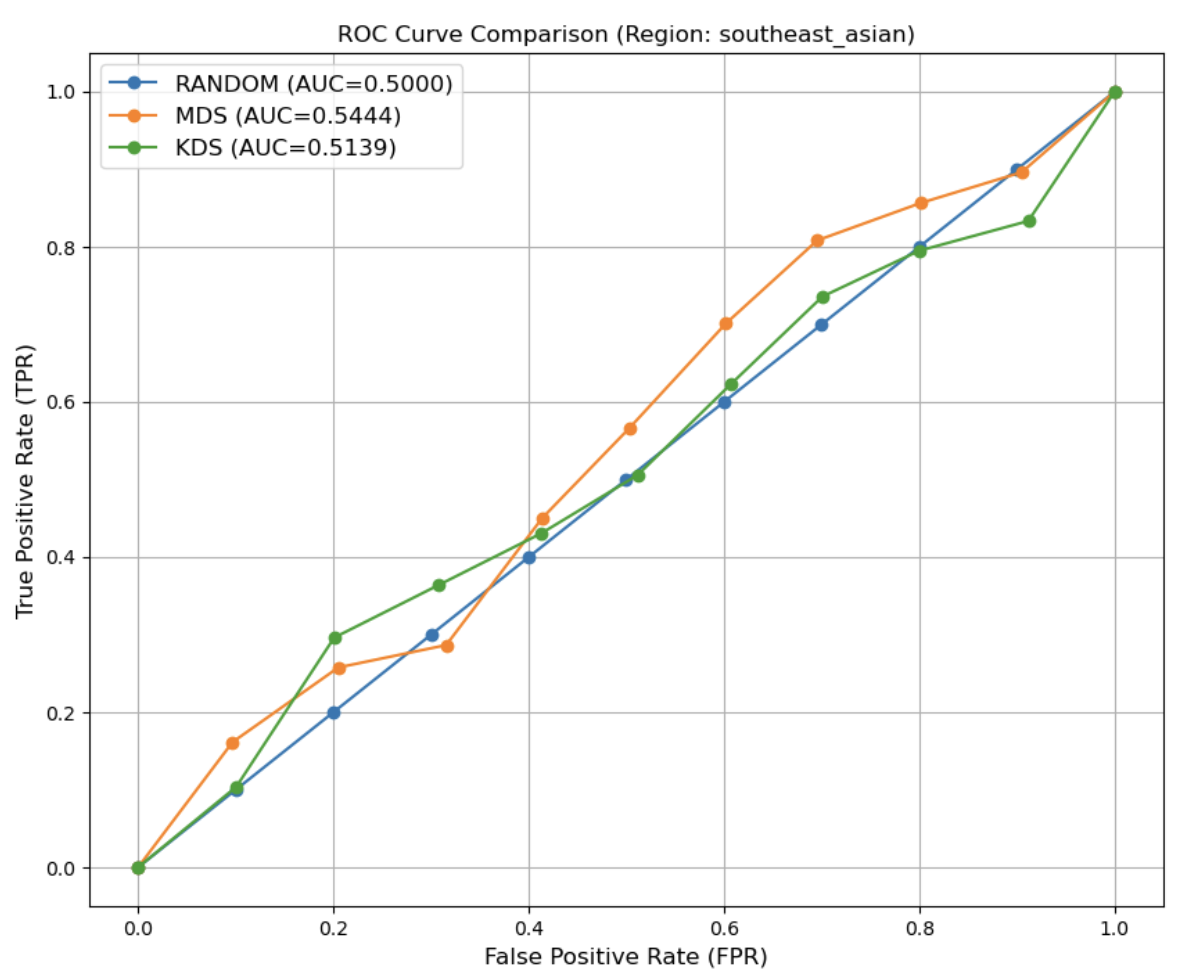}
  \end{minipage}
  \begin{minipage}{0.49\linewidth}
    \includegraphics[width=0.9\linewidth]{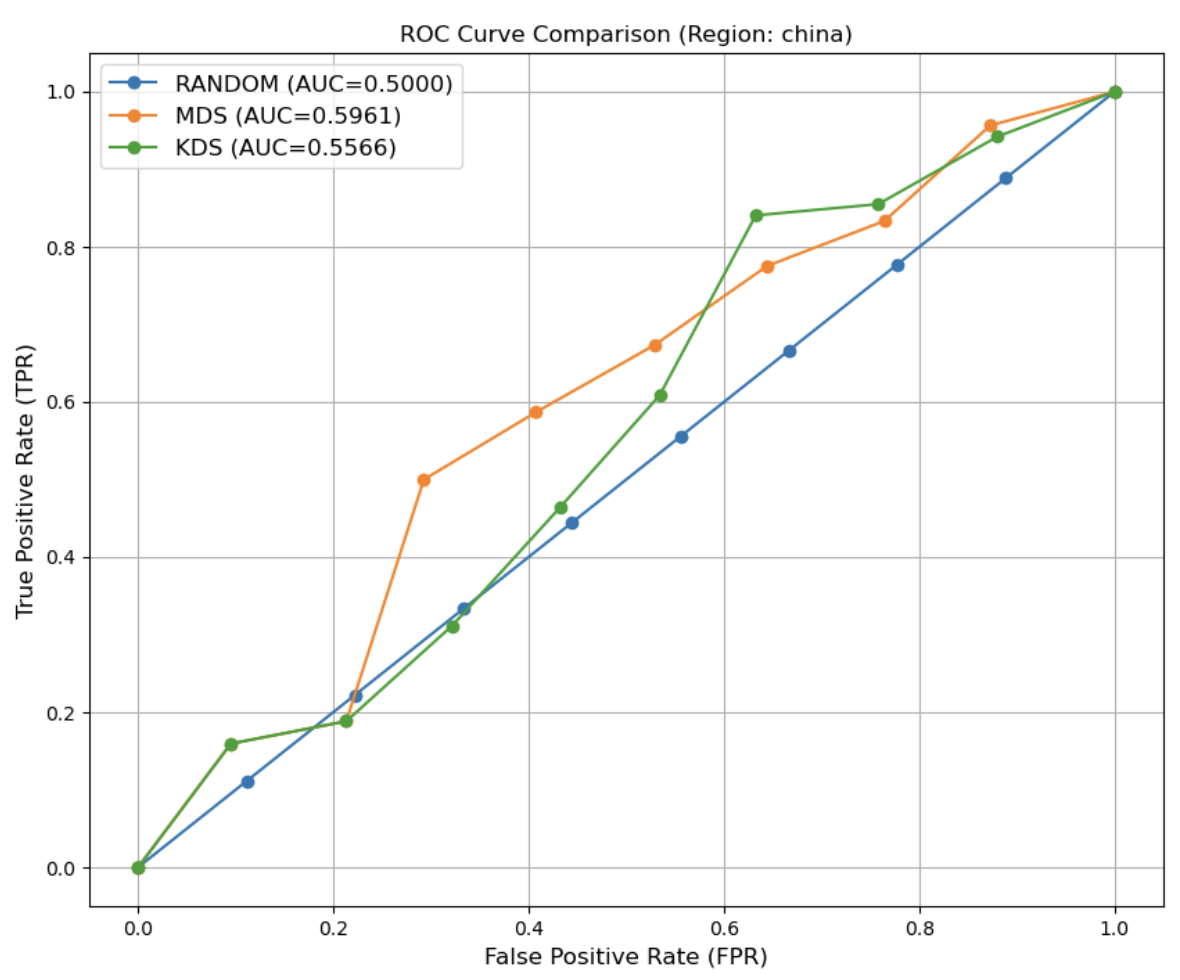}
  \end{minipage}
  \begin{minipage}{0.49\linewidth}
    \includegraphics[width=0.9\linewidth]{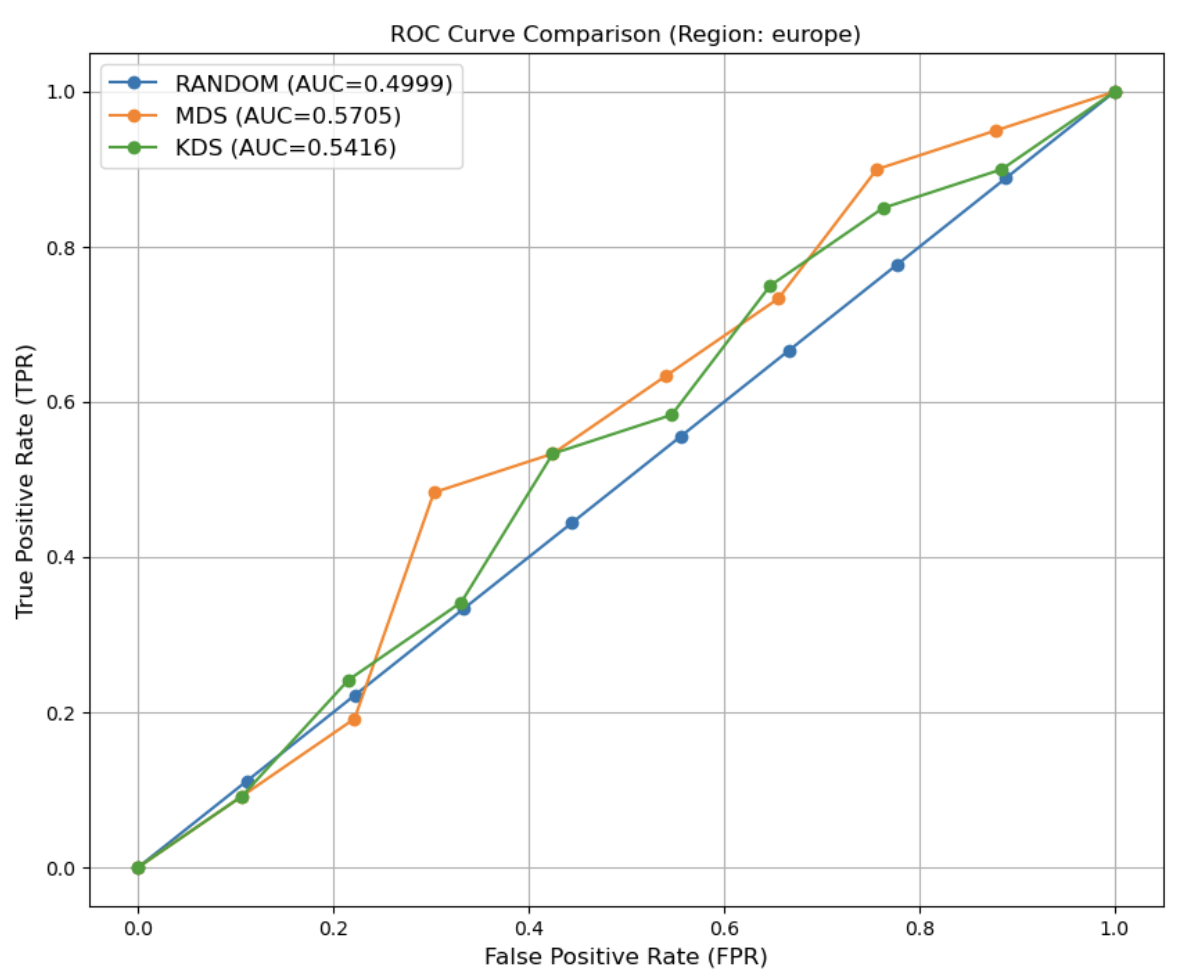}
  \end{minipage}
  \caption{ROC Curve: Taste-Curiosity (Top Left: Average Across All Regions, Top Right: Southeast Asian Food, Bottom Left: Chinese Food, Bottom Right: European Food)}
  \label{roc_curve_taste_curiosity}
\end{figure*}

Figure \ref{roc_curve_taste_curiosity} shows the ROC curve for taste-curiosity.
Looking at the average ROC curve across all regions, the ROC curve of the proposed method based on MDS and KDS is positioned above the Baseline, indicating that the proposed method has higher recommendation performance.
In the comparison between the MDS-based and KDS-based methods, the AUC values were $0.5384$ for the former and $0.5085$ for the latter, demonstrating that, as in Experiment 1, the MDS-based method is superior to the KDS-based method.

Looking at the ROC curves for each region, unlike Experiment 1, the proposed methods outperform the Baseline in all three regions.
However, there was no significantly higher performance than the Baseline, as seen in the ROC curve for Chinese food in Experiment 1.
Additionally, in some metrics such as Precision@3 and Precision@5 in Table \ref{taste_curiosity_ranking_southeast_asia} and Recall@1 in Table \ref{taste_curiosity_ranking_europe}, the proposed methods performed worse than the Baseline, indicating room for improvement.

\vspace{-5mm}
\begin{table}[H]
    \centering
    \caption{Recommendation Performance of Southeast Asian Food}
    \label{taste_curiosity_ranking_southeast_asia}
    \scalebox{0.9}{
      \begin{tabular}{l|llllllllll}
      \hline \hline
      Method & P@1 & P@3 & P@5 & R@1 & R@3 & R@5 & NDCG@1 & NDCG@3 & NDCG@5 \\\hline
      Baseline & 0.1615 & {\bf 0.1615} & {\bf 0.1615} & 0.1000 & 0.2999 & 0.5000 & 0.1615 & 0.2476 & 0.3317 \\
      MDS & {\bf 0.1923} & 0.1282 & {\bf 0.1615} & {\bf 0.1615} & 0.2865 & {\bf 0.5667} & {\bf 0.1923} & 0.2561 & {\bf 0.3766} \\
      KDS & 0.1538 & 0.1538 & 0.1462 & 0.1038 & {\bf 0.3635} & 0.5058 & 0.1538 & {\bf 0.2799} & 0.3415 \\\hline \hline
      \end{tabular}
    }
\end{table}

\vspace{-10mm}
\begin{table}[H]
    \centering
    \caption{Recommendation Performance of Chinese Food}
    \label{taste_curiosity_ranking_china}
    \scalebox{0.9}{
      \begin{tabular}{l|llllllllll}
      \hline \hline
      Method & P@1 & P@3 & P@5 & R@1 & R@3 & R@5 & NDCG@1 & NDCG@3 & NDCG@5 \\\hline
      Baseline & 0.1547 & 0.1547 & 0.1546 & 0.1111 & 0.3335 & 0.5556 & 0.1547 & 0.2565 & 0.3548 \\
      MDS & {\bf 0.2609} & {\bf 0.2464} & {\bf 0.1913} & {\bf 0.1594} & {\bf 0.5000} & {\bf 0.6739} & {\bf 0.2609} & {\bf 0.3710} & {\bf 0.4463} \\
      KDS & {\bf 0.2609} & 0.1739 & 0.1826 & {\bf 0.1594} & 0.3116 & 0.6087 & {\bf 0.2609} & 0.2738 & 0.4037 \\\hline \hline
      \end{tabular}
    }
\end{table}

\vspace{-10mm}
\begin{table}[H]
    \centering
    \caption{Recommendation Performance of European Food}
    \label{taste_curiosity_ranking_europe}
    \scalebox{0.9}{
      \begin{tabular}{l|llllllllll}
      \hline \hline
      Method & P@1 & P@3 & P@5 & R@1 & R@3 & R@5 & NDCG@1 & NDCG@3 & NDCG@5 \\\hline
      Baseline & 0.1277 & 0.1278 & 0.1278 & {\bf 0.1111} & 0.3333 & 0.5555 & 0.1277 & 0.2443 & 0.3380 \\
      MDS & {\bf 0.1500} & {\bf 0.2000} & {\bf 0.1500} & 0.0917 & {\bf 0.4833} & {\bf 0.6333} & {\bf 0.1500} & {\bf 0.3193} & {\bf 0.3795} \\
      KDS & {\bf 0.1500} & 0.1333 & 0.1400 & 0.0917 & 0.3417 & 0.5833 & {\bf 0.1500} & 0.2488 & 0.3560 \\\hline \hline
      \end{tabular}
    }
\end{table}

\vspace{-10mm}
\begin{table}[H]
    \centering
    \caption{Average Recommendation Performance Across Three Regions}
    \label{taste_curiosity_ranking_all_regions}
    \scalebox{0.9}{
      \begin{tabular}{l|llllllllll}
      \hline \hline
      Method & P@1 & P@3 & P@5 & R@1 & R@3 & R@5 & NDCG@1 & NDCG@3 & NDCG@5 \\\hline
      Baseline & 0.1480 & 0.1480 & 0.1480 & 0.1074 & 0.3223 & 0.5370 & 0.1480 & 0.2494 & 0.3415 \\
      MDS & {\bf 0.2011} & {\bf 0.1915} & {\bf 0.1676} & {\bf 0.1375} & {\bf 0.4233} & {\bf 0.6246} & {\bf 0.2011} & {\bf 0.3155} & {\bf 0.4008} \\
      KDS & 0.1882 & 0.1537 & 0.1563 & 0.1183 & 0.3389 & 0.5659 & 0.1882 & 0.2675 & 0.3670 \\\hline \hline
      \end{tabular}
    }
\end{table}

\section{Conclusion}
In this study, we addressed the challenge to recommend food that balance comfort and curiosity.

First, we quantified the qualitative concepts of ``comfort'' and ``curiosity'' that food possess, taking into account past food history data. 
Next, we formulated curiosity (return) per unit risk obtained by eating food using these concepts and proposed a ranking method for food recommendations that considers the balance between comfort and curiosity.

Finally, we used a newly constructed dataset obtained through crowdsourcing, comparing the performance of the proposed methods using KDS and MDS against a baseline of random recommendation.
The results indicated that the MDS-based method outperformed the Baseline when taste was associated with comfort and ingredients with curiosity.
In contrast, no significant difference was observed between the KDS-based method and the Baseline.
Additionally, when taste was associated with curiosity and ingredients with comfort, both the MDS-based and KDS-based methods showed improved performance compared to the Baseline.
Overall, the MDS-based method demonstrated superiority across all cases.

\begin{credits}
\subsubsection{\ackname}
This work was partly supported by JSPS KAKENHI (23K28094)
\end{credits}
%
%
%
\bibliographystyle{splncs04}
\bibliography{arXiv_Yuto_Sakai}

\end{document}